# Valuation of Equity Linked Securities with Guaranteed Return

David Xiao

## ABSTRACT

Equity-linked securities with a guaranteed return become very popular in financial markets ether as investment instruments or life insurance policies. The contract pays off a guaranteed amount plus a payment linked to the performance of a basket of equities averaged over a certain period. This paper presents a new model for valuing equity-linked securities. Our study shows that the security's price can be replicated by the sum of the guaranteed amount plus the price of an Asian style option on the basket. Analytical formulas are derived for the security's price and corresponding hedge ratios. The model appears to be accurate over a wide range of underlying security parameters according to numerical studies. Finally, we use our model to value a segregated fund with a guarantee at maturity.

# 1 Introduction

An Equity-linked security with guaranteed return is a financial instrument that provides investors a guaranteed amount together with equity market upside exposure. The yield from the security depends on the performance of a basket consisting of one or more equities or equity indices. Here the basket level is given by a certain weighted sum of the respective price of each equity or index. The security's return at maturity is based on the arithmetic average of the basket's level over certain points in time, but is floored by a guaranteed amount.

Equity-linked derivatives become popular mainly due to the guaranteed minimum return that the investor receives. This guarantee feature allows investors to benefit from the upside potential of equity growth without full exposure to the downside risk.

The upside potential for this product is unlimited. The potential positive return on the securities is the same as the positive price return on the underlying asset. The downside risk is limited due to a guaranteed return. However, there is an opportunity cost even where an investor receives a guaranteed return in down markets, because investors have lost the use of their invested principal for the term of the contract.

The market today for these products has evolved significantly from traditional single stock equity derivatives to basket-based products. Traditional forms of equity derivatives were generally used to raise equity capital for companies, whereas these equity-linked products are increasingly being used in the process of debt fund raising through new issue arbitrage techniques.

The surge in the sale of equity-linked securities has led to many discussions on the valuation and risk management of these products. Tiong (2000) studies the pricing of the embedded financial option in such contracts with the asset price following a geometric Brownian motion.

Gerber, et al. (2012) carry out the analysis in a risk-neutral framework through Esscher transforms and develop explicit pricing results within that framework.

Henderson & Pearson (2011) argue that the issuers exploit the behavioral bias in investors' mind to sell structured products and analyze US based structured products for returns compared to risk free deposits.

Roger (2008) analyzes capital protected notes for loss averse investors and suggests that the reason for price premium is to account for service provided by the issuers.

Rieger (2012) analyzes the reason for retail investors buying structured products and concludes the probability mis-estimation and behavioral biases play dominant role in investors mind while buying such products.

Deng, et al. (2015) publish a detailed valuation guide for various types of structured products and derive various payoff patterns for structured products.

Zhang et al. (2020) utilize the exponential Lévy process for modeling the stock price process to analyze the equity-linked pricing problem. By using the Fast Fourier Transform, they derive the price of the structured products and obtain the price for various payoffs. Their numerical results are compared to those computed using B-spline functions of different orders to demonstrate the efficiency and accuracy of their proposed algorithm.

Wang et al. (2021) analyze the valuation problem of equity-linked instruments with regime-switching jump diffusion models. Their method of Fourier expansion and Fourier transform has been used to derive closed expressions for some contracts. Their method's effectiveness is demonstrated by numerical values that confirm its efficiency.

Kirkby and Nguyen (2021) focus their work on determining the payoff of equity-linked products and are able to derive a closed form of the price of such products when the risky index process follows the exponential Lévy process

This paper presents a new model for valuing equity-linked security that provides a guaranteed return. This structured product can be divided into two components: a fixed return and an embedded Asian style option on the basket of indices. In this paper we focus on valuing the embedded option and its sensitivity to different market conditions.

We use the model to price a segregated fund whose value at maturity is guaranteed to be greater than the starting invested principal. The fund holder incurs, in addition to a management fee, a protection fee towards the fund's guaranteed minimum value at maturity. We show that the value of this guaranteed return, which amounts to the price of a certain European style put option, can be computed using our model considered above.

We test the model numerically by compared against both a Monte Carlo benchmark and a Levy based approximate option pricing formula. The numerical results show close agreement with the Monte Carlo benchmark over a wide range of option parameters, indicating prima facie that the model is quite accurate. It also shows that our model is more accurate than Levy model.

The rest of this article is organized as follows: First we elaborate the form of the equity-linked security with guaranteed return. Second, we present a new model for valuing this security. Third, we show numerical pricing results. Next, we describe a type of segregated fund, and how to value its maturity guarantee using the model described. Finally, the conclusions are provided.

## 2    Product Definition

An equity-linked security is a financial product that differs from a standard fixed-income security in that the payoff is based on the performance of a single stock, basket of stocks or equity index. The final payment at maturity is determined by the appreciation of the underlying equity.

This type of instrument is appropriate for conservative equity investors or fixed income investors who desire equity exposure with controlled risk. We can define the product more rigorously as follows:

We consider a security whose payoff depends on the return from a finite number, $M$, of equity-linked indices. Let $I_t^j$, for $j = 1,\ldots,M$, denote the price of the $j^{th}$ index at time equal to $t$, and let $\omega_j$ denote a fixed, positive weight corresponding to this index. Next let $T$ denote the security's maturity. Furthermore let $\{t_1,\ldots,t_N\}$, where $N > 0$ and $0 < t_1 < \ldots < t_N \leq T$, be a finite set of observation times. Finally let $P$ denote an amount guaranteed to the security holder at maturity.

The payoff at maturity depends on the weighted sum, over each index, of the relative change in the arithmetic average of the index's price, with respect to the set observation points above, from the index's initial level. Formally the payoff at maturity is given by

$$\max\left(P + \sum_{i=1}^{M}\omega_i \frac{\frac{1}{N}\sum_{j=1}^{N}I_{t_j}^i - I_0^i}{I_0^i},\ P\right) = P + \max\left(\sum_{i=1}^{M}\omega_i \frac{\frac{1}{N}\sum_{j=1}^{N}I_{t_j}^i - I_0^i}{I_0^i},\ 0\right). \qquad (2.1)$$

Next let

$$Z_t = \sum_{j=1}^{M}\alpha_j I_t^j$$

denote the price at time $t$ of a basket of the equity-linked indices above; here $\alpha_j = \dfrac{\omega_j}{I_0^j}$ is the ratio of the $j^{th}$ index's weight over the index's initial level.

Then the payoff (2.1) is equivalent to

$$P + \max\left(\frac{1}{N}\sum_{i=1}^{N} Z_{t_i} - \sum_{i=1}^{M} \omega_i, 0\right),$$

which is the sum of the payoff from an Asian style option sampled at the discrete points above plus the guaranteed component.

## 3  Model Description

In this section we present our model for pricing an Asian style option with payoff at maturity, $T$, of the form

$$\max\left(\frac{1}{N}\sum_{i=1}^{N} Z_{t_i} - \sum_{i=1}^{M} \omega_i, 0\right). \tag{3.1}$$

Here we assume that each index follows geometric Brownian motion with drift under its respective risk-neutral probability measure. Each index is then expressed under the domestic risk-neutral probability measure by a corresponding change of measure. Observe that, under these assumptions, the random variable

$$Y = \frac{1}{N}\sum_{i=1}^{N} Z_{t_i} \tag{3.2}$$

is not log-normally distributed. This, then, makes it mathematically difficult to value the payoff (3.1) using analytical techniques.

The standard Levy approach (see Levy (1992)) towards valuing the payoff (3.1) is to approximate Y in (3.2) by a log-normally distributed random variable. Here the defining parameters for the log-normal random variable are uniquely determined by matching its first two moments with those of Y.

The option value is then given from an analytical formula by taking the expected value of the payoff (3.1), but where the underlying security value, Y, is replaced by that of the log-normally distributed random variable. As we will see in Section 4, however, this method may produce largely inaccurate pricing and hedging results depending on option's tenor and volatility parameters.

Our valuation approach aims to match more moments, and can be viewed as an extension of Levy's. Specifically, we approximate $Y$ in (3.2) by a shifted log-normal random variable, of the form

$$a + e^{b+c\varepsilon}, \qquad (3.3)$$

where $a$ and b are constants, $c$ is a positive constant, and $\varepsilon$ is a standard, normally distributed random variable. Here $a$, $b$ and $c$ are uniquely determined, with analytical form, by matching the first three moments of Y with those of (3.3). An analytical, approximate option pricing formula is then derived by taking the expected value of the payoff (3.1), but where the underlying security's value is replaced by that of the shifted log-normal random variable.

Assume that, under the domestic risk neutral probability measure, the process $\{I_t^j \mid t \geq 0\}$, for $j = 1,...,M$, satisfies a stochastic differential equation of the form

$$dI_t^j = I_t^j \left( \mu_j dt + \sigma_j dW_t^j \right)$$

where $\mu_j$ is a constant drift parameter, $\sigma_j$ is a constant volatility parameter, and $\{W_t^j | t \geq 0\}$ is a standard Brownian motion.

Suppose also that the Brownian motions $\{W_t^j | t \geq 0\}$ and $\{W_t^k | t \geq 0\}$, for $j, k \in \{1,..., M\}$, have a constant instantaneous correlation coefficient, $\rho_{jk}$. The first moment of $Y$ then equals

$$E(Y) = \frac{1}{N} \sum_{k=1}^{N} \sum_{i=1}^{M} \alpha_i I_0^i e^{\mu_i t_k}.$$

Furthermore, the second moment of $Y$ is given by

$$E(Y^2) = \frac{1}{N^2} \sum_{k,l \in \{1,...,N\}} E(Z_{t_k} Z_{t_l}),$$

where

$$E(Z_{t_k} Z_{t_l}) = \sum_{m,n \in \{1,...,M\}} \alpha_m \alpha_n I_0^m I_0^n e^{\mu_m t_k + \mu_n t_l + \sigma_m \sigma_n \rho_{mn} \min(t_k, t_l)}.$$

Also, the third moment of $Y$ equals

$$E(Y^3) = \frac{1}{N^3} \sum_{l,m,n \in \{1,...,N\}} E(Z_{t_l} Z_{t_m} Z_{t_n}),$$

where

$$E(Z_{t_l} Z_{t_m} Z_{t_n}) = \sum_{i,j,k \in \{1,...,M\}} \alpha_i \alpha_j \alpha_k I_0^i I_0^j I_0^k e^{\mu_i t_l + \mu_j t_m + \mu_k t_n + \sigma_i \sigma_j \rho_{ij} \min(t_l, t_m) + \sigma_i \sigma_k \rho_{ik} \min(t_l, t_n) + \sigma_j \sigma_k \rho_{jk} \min(t_m, t_n)}.$$

By matching the first three moments of $Y$ with those of the shifted log-normal random variable (3.3), we obtain the system of nonlinear equations

$$E(Y) = a + e^{b+\frac{c^2}{2}}, \tag{3.4a}$$

$$E(Y^2) = a^2 + 2ae^{b+\frac{c^2}{2}} + e^{2b+2c^2}, \tag{3.4b}$$

$$E(Y^3) = a^3 + 3a^2 e^{b+\frac{c^2}{2}} + 3ae^{2b+2c^2} + e^{3b+\frac{9}{2}c^2}, \tag{3.4c}$$

for the unknowns $a$, $b$ and $c$. It can be shown that, under certain conditions, the nonlinear system of equations above has a closed-form, unique, real solution.

Let $r$ denote the risk-free interest rate (see https://finpricing.com/lib/IrCurve.html) for a term equal to the option maturity, $T$. The Asian style option with payoff (3.1) then has value

$$\Omega = e^{-rT} E\left( \max\left( \frac{1}{N} \sum_{i=1}^{N} Z_{t_i} - \sum_{i=1}^{M} \omega_i, 0 \right) \right), \tag{3.7a}$$

which we approximate by

$$\tilde{\Omega} = e^{-rT} E\left( \max\left( a + e^{b+c\varepsilon} - \sum_{i=1}^{M} \omega_i, 0 \right) \right). \tag{3.7b}$$

Let $X$ denote $\sum_{i=1}^{M} \omega_i$. If $X > a$, then (3.7b) equals

$$e^{-rT} \int_{\frac{\log(X-a)-b}{c}}^{+\infty} \left(a + e^{b+cy} - X\right) f(y) dy$$

$$= e^{-rT}\left[(a-X)n\left(\frac{b-\log(X-a)}{c}\right) + e^{b+\frac{c^2}{2}} n\left(c + \frac{b-\log(X-a)}{c}\right)\right],$$

where $f(y) = \frac{1}{\sqrt{2\pi}} e^{-\frac{1}{2}y^2}$ is the probability density function for a standard, normally distributed random variable and $n(z) = \int_{-\infty}^{z} f(y) dy$ is the corresponding cumulative distribution function.

If $X \leq a$, then (3.7b) equals

$$e^{-rT}\left(a - X + e^{b+\frac{c^2}{2}}\right).$$

We may be interested in the option delta, $\frac{\partial \Omega}{\partial I_0^j}$, and the option Vega, $\frac{\partial \Omega}{\partial \sigma_j}$, for $j = 1,\ldots,M$. These hedge ratios are respectively approximated by $\frac{\partial \tilde{\Omega}}{\partial I_0^j}$ and $\frac{\partial \tilde{\Omega}}{\partial \sigma_j}$, for $j = 1,\ldots,M$, and are obtained from direct differentiation of $\tilde{\Omega}$ using the chain rule.

Let the floor level, $F$, be given such that $F \leq -1$. Furthermore let $N = 0$ be the number of price returns to be capped. The combined price return is then of the form

$$\sum_{i=1}^{M} \max\left(\frac{I_i}{I_{i-1}} - 1, F\right) = \sum_{i=1}^{M}\left(\frac{I_i}{I_{i-1}} - 1\right),$$

$$= -M + \sum_{i=1}^{M} \frac{I_i}{I_{i-1}},$$

$$= -M + \sum_{i=1}^{M} e^{\left(r-q-\frac{\sigma^2}{2}\right)(t_i - t_{i-1}) + \sigma(W_{t_i} - W_{t_{i-1}})},$$

since $\frac{I_i}{I_{i-1}} > 0$, for $i \in \{1,..,M\}$. The value of the combined price return then equals

$$E\left(\frac{-M + \sum_{i=1}^{M} e^{\left(r-q-\frac{\sigma^2}{2}\right)(t_i - t_{i-1}) + \sigma(W_{t_i} - W_{t_{i-1}})}}{e^{rT_M}}\right) = e^{-rT_M}\left(-M + \sum_{i=1}^{M} e^{(r-q)(t_i - t_{i-1})}\right).$$

## 4  Numerical Results

It is interesting to compare the accuracy of our approximate option pricing formula, as well as that of a Levy based pricing formula, against a Monte Carlo benchmark. To this end we consider the following Levy based pricing approach. Let $U$ be a log-normally distributed random variable, of the form

$$U = e^{a+b\varepsilon},$$

where $a$ is a constant, $b$ is a positive constant, and $\varepsilon$ is a standard, normally distributed random variable. We choose $a$ and $b$ by matching the first two moments of the basket's price at maturity with those of the log-normal random variable $U$. We then approximate the option's price, (3.7a), by

$$e^{-rT} E\left(\max\left(U - \sum_{i=1}^{M} \omega_i, 0\right)\right).$$

We have implemented both our pricing model, described in Section 3, and the Levy based approach above.

As an example, we consider the Asian style option arising from a security dependent on the return from a basket of five indices. Here the payoff, of the form (3.1), depends on the arithmetic average of the basket's price at twelve observation points. These points are respectively set to the last business day in each of the eleven months that precede the month in which the security matures and the business day that immediately precedes the maturity date.

Here the security was issued on July 23, 2018, and matures on July 23, 2023; the valuation date is on June 2, 2019. In Figure 4.2 we show the initial level and corresponding weight for each index, as well as the observation points.

The valuation of structured products depends on implied volatility that has a typical "skewed" shape shown in Figure 4.1:

**Figure 4.1.** Implied volatility surface with skew.

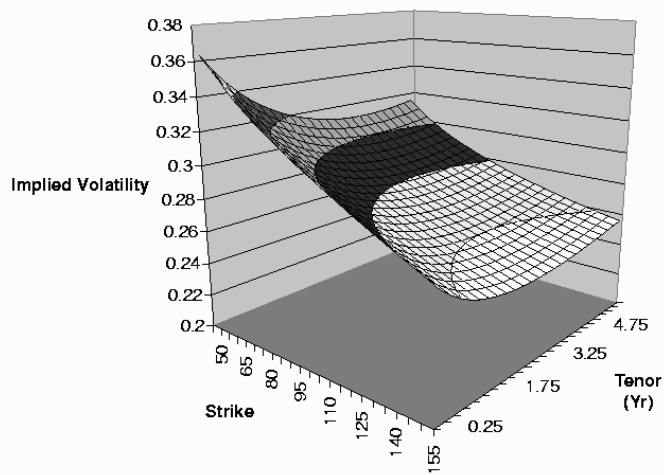

**Figure 4.2.** Snapshot of basket description screen (here we display two significant digits).

| Basket Components | | | | | |
|---|---|---|---|---|---|
| Index | 1 | 2 | 3 | 4 | 5 |
| Initial level | 2421.04 | 391.64 | 1147.27 | 15944.36 | 2913.59 |
| Weight | 25 | 30 | 10 | 2.5 | 2.5 |

| Observation Schedule | | |
|---|---|---|
| 1 | 3.25 | 2022-08-31 |
| 2 | 3.33 | 2022-09-30 |
| 3 | 3.41 | 2022-10-31 |
| 4 | 3.50 | 2022-11-30 |
| 5 | 3.58 | 2022-12-31 |
| 6 | 3.67 | 2023-01-31 |
| 7 | 3.74 | 2023-02-28 |
| 8 | 3.83 | 2023-03-31 |
| 9 | 3.91 | 2023-04-30 |
| 10 | 3.99 | 2023-05-31 |
| 11 | 4.08 | 2023-06-30 |
| 12 | 4.14 | 2023-07-22 |

In Figure 4.3 we show pricing results for various embedded options specified from the parameters in Figure 4.2. Here the Volatility Shift parameter indicates a respective relative shift to all original volatility parameter values.

The corresponding benchmark option prices, shown in Figure 4.3, are numerical values for Formula (3.7a), which were computed using crude Monte Carlo simulation based on four million sample paths with .7% standard error. We note that, for the case of zero shift to the volatilities, the parameter values for the shifted log-normal random variable,

$$a + e^{b+c\varepsilon},$$

were computed as $a = .3259$, $b = -.6821$ and $c = 0.3332$.

**Figure 4.3.** Numerical option pricing results (expressed as a percentage of the notional amount).

| Volatility shift | Model price | MC Price | Levy Price |
| --- | --- | --- | --- |
| -50 | 12.59 | 12.61 | 12.61 |
| 0 | 13.48 | 13.50 | 13.74 |
| 50 | 15.10 | 15.13 | 15.80 |
| 100 | 16.99 | 17.06 | 18.37 |

In Figure 4.4 we display various hedge ratios, with respect to the first index, for the option specified from the original parameters shown in Figure 4.2. The hedge ratios based on Formula (3.7b) are from the direct, analytical differentiation of (3.7b). Benchmark hedge ratios are computed using a one-sided finite difference approximation applied to the true option pricing formula, (3.7a); here numerical values for (3.7a) were obtained using crude Monte Carlo simulation based on 4 million sample paths.

Levy based hedge ratios are based on a finite difference approximation. Observe that the relative error in the Levy based vega value from the Monte Carlo (MC) benchmark is approximately 37%, while the vega from Formula (3.7b) differs from the benchmark only by 5.8%.

**Figure 4.4.** Hedge ratios with respect to the first index.

| Hedge ratio | Model value | MC value | Levy value |
|---|---|---|---|
| Price Delta | 0.0083 | 0.0083 | 0.0082 |
| Vega | 2.76 | 2.94 | 1.85 |

## 5 Segregated Fund Valuation

We consider a type of segregated fund that invests in various foreign and domestic equities and bonds. We assume that the fund provides a maturity guarantee, that is, the fund's price at maturity is assured to be greater than the original invested amount. We also assume that the fund has no dynamic lapse or reset features, and that the holder pays periodic management and protection fees.

We model the fund's value by the price of basket of representative equity and bond-linked indices; the maturity guarantee then measures the net shortfall from the basket's constituent indices. Specifically suppose that the basket contains a fixed number, $N$, of indices. Furthermore let $I_t^i$, for $i = 1,...,N$, denote the price of the $i^{th}$ index at time $t$.

Next let $P$ denote the principal amount originally invested in the fund. Assume also that at the fund's outset a percentage, $v_i$, of the principal is invested in the $i^{th}$ $(i = 1,...,N)$ index; the initial number of units, $u_i$, associated with the $i^{th}$ index then equal

$$u_i = \frac{v_i P}{I_0^i}.$$

Let $T$ denote the fund's maturity. Assume that protection and management fees are both collected at a set of times, $\{t_1,..., t_M\}$, where $0 < t_1 < ... < t_M < T$. Suppose also that the protection and management fee are taken, at time $t_i$ $(i = 1,...,M)$, as respective

percentages, $p_i$ and $m_i$, of the fund's price at $t_i$. The fund's price at maturity then equals

$$\sum_{i=1}^{N} \omega_i I_T^i \qquad (5.1)$$

where $\omega_i = u_i \prod_{j=1}^{M} [1 - (m_j + p_j)]$.

Suppose that the invested principal, $P$, is 100% guaranteed at maturity. The payoff at maturity from this guarantee then equals

$$\max\left(P - \sum_{i=1}^{N} \omega_i I_T^i, 0\right), \qquad (5.2)$$

which has the same form as that of a European style put option. Our approach towards valuing the payoff above is based on that presented in Section 3.

Specifically, we assume that the $i^{th}$ ($i = 1,...,N$) index's price process, $\{I_t^i \mid t > 0\}$, follows geometric Brownian motion with drift under the domestic risk-neutral probability measure. We then approximate the basket's price at maturity, $\sum_{i=1}^{N} \omega_i I_T^i$, by a shifted log-normal random variable, of the form

$$a + e^{b+c\varepsilon}. \qquad (5.3)$$

Here the parameters $a$, $b$ and $c$ are uniquely determined, as described in Section 3, by matching the first three moments of the shifted log-normal random variable with those of the basket's price at maturity. We next approximate the payoff (5.2) by replacing the basket's value at maturity with that of the shifted log-normal random variable, that is,

$$\max\left(P-\left(a+e^{b+c\varepsilon}\right),0\right). \tag{5.4}$$

Let $r$ denote the constant risk-free rate for a period equal to the fund's maturity, $T$. The payoff (5.4) then has value

$$e^{-rT} E\left(\max\left(P-\left(a+e^{b+c\varepsilon}\right),0\right)\right) \tag{5.5}$$

where $E$ denotes the domestic risk-neutral probability measure. If $a < P$, then (5.5) equals

$$e^{-rT}\left((P-a)n\left(\frac{\log(P-a)-b}{c}\right) - e^{b+\frac{c^2}{2}} n\left(\frac{\log(P-a)-b}{c}-c\right)\right)$$

where $n$ is the cumulative distribution function for a standard, normally distributed random variable. If $a \geq P$, then (5.5) equals zero.

We note that our analytical model above for valuing the European style put option provides for a significant speed-up over alternative Monte Carlo or quasi-Monte Carlo pricing methods.

## 6 Summary

Most forms of equity linked structured products have their performance linked to an underlying benchmark such as an equity index, real estate index, commodities, interest rates etc.

We consider a type of equity-linked security that provides a guaranteed return. We show that the security's price is given by a guaranteed component plus the value of an embedded Asian style option on a basket of equity-linked indices.

This paper presents a new model for valuing the embedded option. Our approach towards pricing the embedded option is to approximate the option's underlying security value using a shifted log-normal random variable. Here the defining parameters for this random variable are given from the analytical, unique solution to a system of non-linear equations arising from a moment matching technique.

An analytical, approximate option pricing formula is then derived by taking the expected value of the payoff and modelling the underlying security's value as the shifted log-normal random variable.

This analytic, approximate pricing model is numerically compared against both a Monte Carlo benchmark and a Levy based approximate option pricing formula. Our pricing formula shows close agreement with the Monte Carlo benchmark over a wide range of option parameter values. The Levy model, however, shows much larger relative errors (e.g., as much as 8%) depending on the option tenor and volatility parameter values.

We also show that our model provides analytical formulas for hedge ratios, from the direct differentiation of the approximate option pricing formula. These formulas are numerically compared against benchmark Monte Carlo based hedge ratios. Numerical results indicate that Delta hedge ratios are in close agreement with the corresponding benchmark, but that the Vega hedge ratios differ by approximately 6% under average parameter values. Hedge ratios computed from the Levy approximation, however, show much larger errors (e.g., as large as 37%) depending on the tenor and volatility.

We also consider a type of segregated fund with a guarantee at maturity. We demonstrate that the value of the guaranteed return is modeled as the price of a certain European style put option on a basket of indices; furthermore, this price can be computed using the model described in Section 3.

In summary, this article presents a new model for the analytical, accurate pricing of certain Asian style options on a basket of underlying equity-linked indices. Furthermore, the model can be applied to value the embedded European style put option arising from a certain type of segregated fund. Although our analytical valuation model provides for a significant speed-up over an alternative Monte-Carlo pricing method, the accuracy of the model degrades depending on the tenor and the underlying volatility parameter values.